# Super-Resolution and Reconstruction of Sparse Sub-Wavelength Images

**Snir Gazit,[1] Alexander Szameit,[1] Yonina C. Eldar,[2] and Mordechai Segev[1]**

1. *Department of Physics and Solid State Institute, Technion, Haifa 32000, Israel*

2. *Department of Electrical Engineering, Technion, Haifa 32000, Israel*

**Abstract**: We use compressed sensing to demonstrate theoretically the reconstruction of sub-wavelength features from measured far-field, and provide experimental proof-of-concept. The methods can be applied to non-optical microscopes, provided the information is sparse.

A fundamental restriction of optical imaging is given by the diffraction limit, stating that the maximal recoverable resolution is set to half of the optical wavelength $\lambda$. This is a direct result of the evanescent nature of all plane-waves associated with spatial frequencies exceeding $1/\lambda$ [1]. Consequently, spatial frequencies higher than $1/\lambda$ are lost, even after short propagation distances of just a few wavelengths. Hence, using optical means to resolve sub-wavelength features from the far-field is virtually impossible. Reaching beyond the sub-wavelength barrier is a subject of intense research. A most useful approach is the Scanning Near-field Optical Microscope (SNOM) [2] which probes the EM field adjacent to the illuminated sample in the "near field" zone. Although the SNOM became a widely used method, its major draw back is the need to scan the sample point by point, preventing its use from real time applications. Alternatively, using the "hyperlens" made of negative-index metamaterials can transform the evanescent modes into propagating ones, enabling direct imaging of sub-wavelength information [3]. However, albeit offering a great promise, negative-index materials are currently severely restricted by high material loss, stringent fabrication requirements and the need to position them in the near-field of the sample. Distributing smaller-than-wavelength fluorescent particles on the sample, exciting them in various (linear and nonlinear) means, repeating the experiments multiple times and ensemble-averaging, constitutes another approach. But this method is not real-time either [4]. A more recent idea employs super-oscillations for sub-wavelength imaging, but this method still requires scanning, either in the near-field or in the plane where the super-oscillations are generated [5]. Apart from these "hardware solutions", several attempts have been made to extrapolate the frequency content above the cut-off frequency dictated by the diffraction limit. However, all of these extrapolation methods are extremely sensitive to noise in the measured data and the assumptions made on the prior knowledge on the information. As such, they have all failed in recovering optical sub-wavelength information [1].

Here, we show that sub-wavelength information can be recovered from the far-field of an optical image, with the only prior knowledge being that the image is sparse. The idea is based on recent compressed sensing (CS) techniques [6], which are generically used for efficient sampling of data. These methods are extremely robust to noise in the measured data. Their only condition is that the information is sparse. However, sparse optical images are common in nature, e.g., living cells, etc. We show theoretically the recovery of sub-wavelength structures from measured data restricted to the low spatial frequency content, and provide experimental proof-of-concept, recovering delicate features (amplitude and phase) that were cut off by spatial filtering.

The underlying logic is that sparsely represented signals hold a very limited number of degrees of freedom, since only a small fraction of their coefficients (in the particular basis representation in which the signal is sparse) are non-zero. This enables to separate two subspaces of the basis functions: the one carrying information of our signal e while the other carries almost none. The aim of our approach is to automatically identify the first subspace and ignore the second. A key observation is that the reconstruction process corresponds to reconstructing a signal from the limited set of measurements from the low spatial frequencies. Hence, we need to compensate for the lost high spatial frequencies by assuming additional prior information on the signal, which is its sparsity. We define $\beta$ as the sparsity level: the relative fraction of non-zero elements in the sparsifying basis of the signal. Since each non-zero component possesses two degrees of freedom – one for its location and the second for its amplitude, one should perform at least a $2\beta$ fraction of the total number of possible measurements, in order to reconstruct the signal. To gain more intuition in which basis the measurements should be preformed, let us consider measurements performed in the same basis in which the signal is sparse. Then, the vast majority of the measurements would be zero and cannot provide information about the true signal. In fact we would have to carry out almost all measurements in that basis in order to ensure exact reconstruction. Instead, we wish to choose the measurement basis such that each measurement of any projection contains information about the signal. This can be achieved by requiring that each measurement basis function has low correlation with each signal basis function. A highly uncorrelated pair of bases obeys an uncertainty principle, preventing a signal from being sparse in both bases and ensuring that, if the signal is sparse in one of the bases, it will be very spread in the other. Therefore almost each projection will yield a non-zero informative measurement in the non-sparse basis. Examples of maximally uncorrelated bases are the spatial and Fourier domains: a highly sparse signal (a single delta function) is Fourier-transformed into a spread function covering the entire spectrum. In our sub-wavelength optical setting, we are restricted to measuring only low frequencies. Accordingly, these will be sufficient to recover the signal if it is sparse in a real-space basis that is uncorrelated with the Fourier basis.

Compressed sensing literature offers efficient recovery algorithms, among them the basis-pursuit method involving minimization of the $l_1$ norm [7]. For optical imaging, an important feature is the ability to detect signals with non-uniform phase. The standard basis-pursuit approach is unable to resolve fine details with different phases, hence we develop an iterative nonlocal thresholding step, facilitating the recovery of signals with phase.

We test the ideas theoretically and are able to reconstruct sub-wavelength information, amplitude and phase. In addition, we devised an experimental proof-of-concept: a standard 4f imaging system with a tunable spatial low pass filter positioned at the Fourier plane. The spatial filter mimics the optical transfer function by eliminating all frequencies above its cut-off. The reconstructed images contain frequency content which significantly exceeds that highest available frequency content. Fig. 1 exhibits the reconstruction of an optical image consisting of three stripes from a single broad strip. Clearly, the idea works well, under conditions physically similar to sub-wavelength imaging: we are able to reconstruct images way beyond cutoff spatial frequencies. We are now working to demonstrate the idea experimentally with true sub-wavelength images of sparse information.

[1] J.W. Goodman. Introduction to Fourier optics. 3rd ed. Englewood, CO: Roberts & Co. Publishers, 2005.
[2] E.A. Ash and G. Nicholls. Super-resolution Aperture Scanning Microscope. *Nature* **237**, 510 (1972).
[3] Z. Jacob, LV. Alexeyev, and E. Narimanov. Optical hyperlens: far-field imaging beyond the diffraction limit. *Opt. Exp.* **14**, 8247 (2006); Z. Liu, H. Lee, Y. Xiong, C. Sun, and X. Zhang. Far-field optical hyperlens magnifying sub-diffraction-limited objects. *Science* **315**, 1686 (2007); I.I. Smolyaninov, Y.J. Hung, and C.C. Davis. Magnifying Superlens in the Visible Frequency Range. *Science* **315**, 1699 (2007).
[4] A. Yildiz, et al.. Myosin V Walks Hand-Over-Hand: Single Fluorophore Imaging with 1.5-nm Localization. *Science* **300**, 2061 (2003).
[5] F.M. Huang and N.I. Zheludev. Super-resolution without evanescent waves. *Nano Lett.* **9**, 1249 (2009).
[6] E.J. Candes and T. Tao. Near-optimal signal recovery from random projections: Universal encoding strategies? *IEEE Transactions on Information Theory* **52**, 5406 (2006).
[7] S.S. Chen, D.L. Donoho, and M.A. Saunders. Atomic Decomposition by Basis Pursuit. *SIAM J. Sci. Comput.* **20**, 33 (1998).

**Fig. 1: Experimental proof-of-principle to retrieve spatial frequencies beyond the cutoff**
(a,b,c) The original information consisting of three vertical stripes (a), its Fourier spectrum (b), and a horizontal cross-section of the amplitude, taken through the real-space information (c).
(d,e,f) Using a slit, the signal is low-pass filtered at the vertical red lines, yielding a highly blurred image (d). The Fourier spectrum now contains now only the lowest frequencies (e), causing the mergence of the three stripes (in real-space) into one, as seen in the horizontal cross section (f).
(g,h,k) Reconstruction using CS methods yields a high quality recovered image (g) and its respective Fourier spectrum (h). The strong correspondence between original and recovery is clearly visible in the horizontal cross section (k).

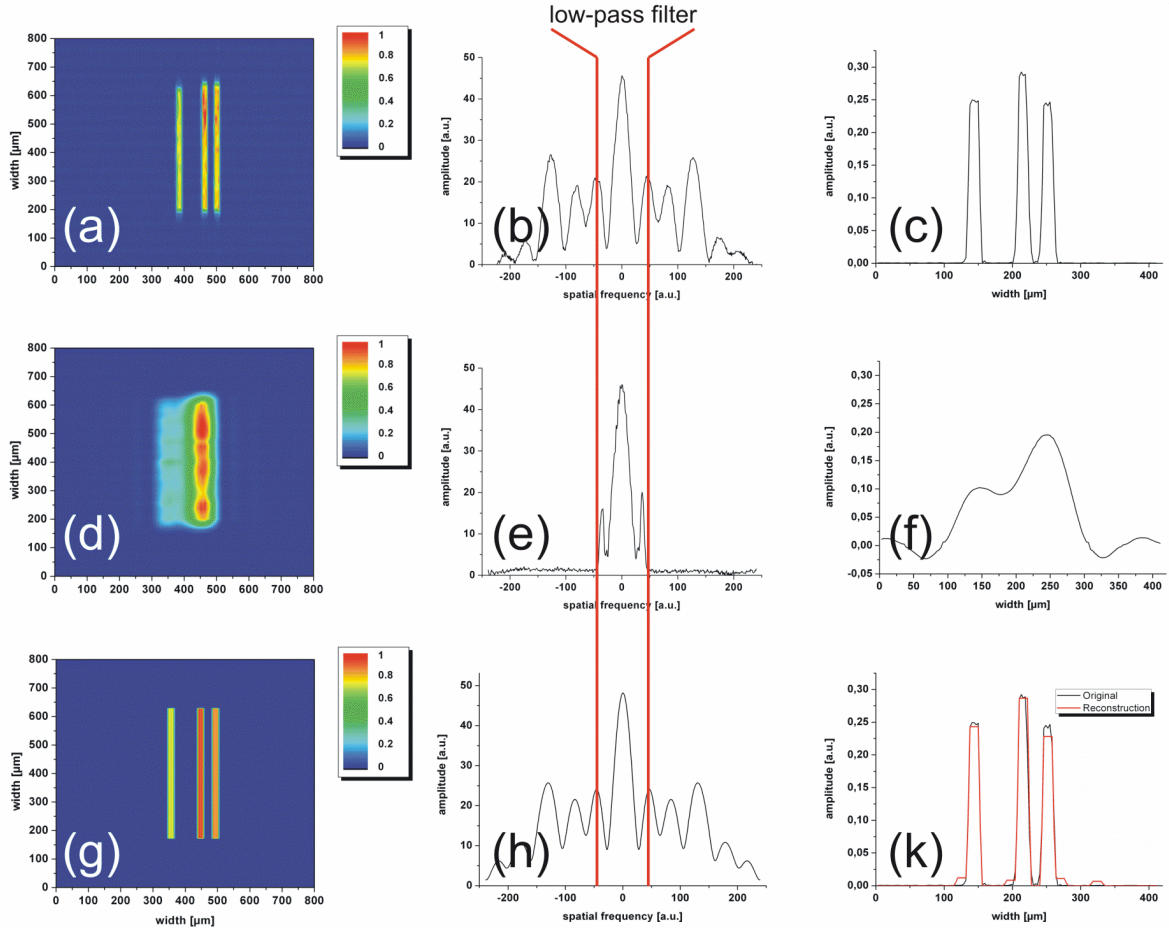

**Fig. 2: Theoretical reconstruction of one-dimensional sub-wavelength information (amplitude and phase).**
(a) The original function, which we want to reconstruct.
(b) The Fourier (plane-wave) spectrum of the original information shown in (a). The vertical red lines indicate the width of the low-pass filter, which for sub-wavelength information is $2/\lambda$.
(c) The distorted image obtained by an inverse Fourier transform on the filtered spectrum; the features are highly blurred.
(d) The low-pass-filtered spectrum; a large fraction of the frequency contents is lost.
(e) Reconstructed image (e) and its spectrum (f) using CS-methods based on the sparsity of the original information. The function is reconstructed perfectly in both real space and Fourier space, including the phase information. Our algorithm is robust against noise.
(g) Adding 1% noise to the filtered spectrum (not shown here), we are still able to reconstruct the original information at high quality in both real space (g) and Fourier space (h).
Amplitude and intensity are given in arbitrary units (a.u.), because the system does not depend on the light intensity

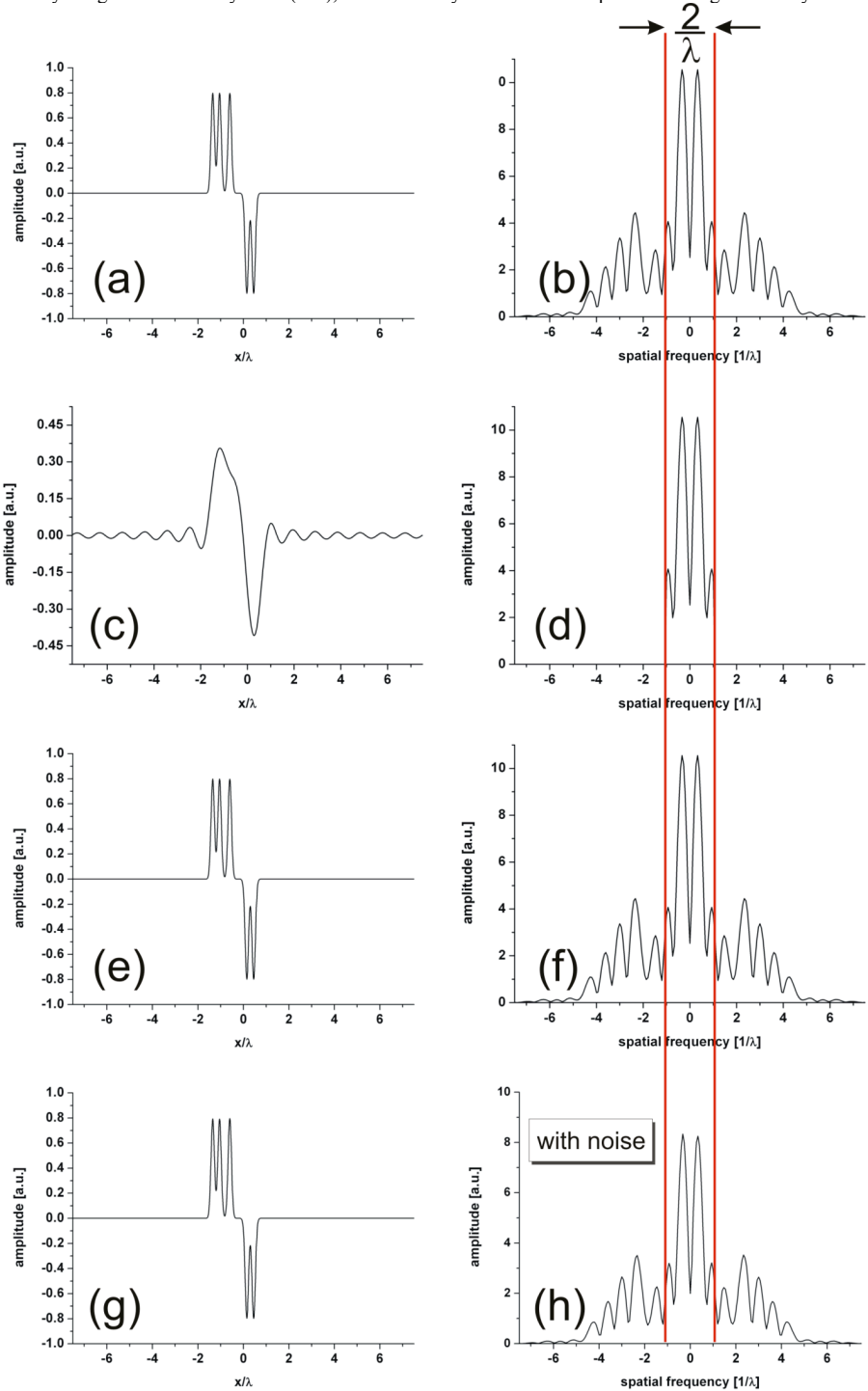

**Fig. 3. Theoretical reconstruction of two-dimensional sub-wavelength information. The information (honeycomb lattice) corresponds to the atomic structure of graphene.**
(a,b) The original information consists of an arrangement of circles, forming the Star of David (a), and its respective Fourier transform (b).
(c,d) After some propagation distance, all spatial frequencies above 1/λ are lost (d), so that the actual observed image is strongly blurred and the fine features cannot be resolved (c).
(e,f) Applying our algorithm reveals the underlying sub-wavelength structure in the real space (e), since the Fourier spectrum is fully restored (f).

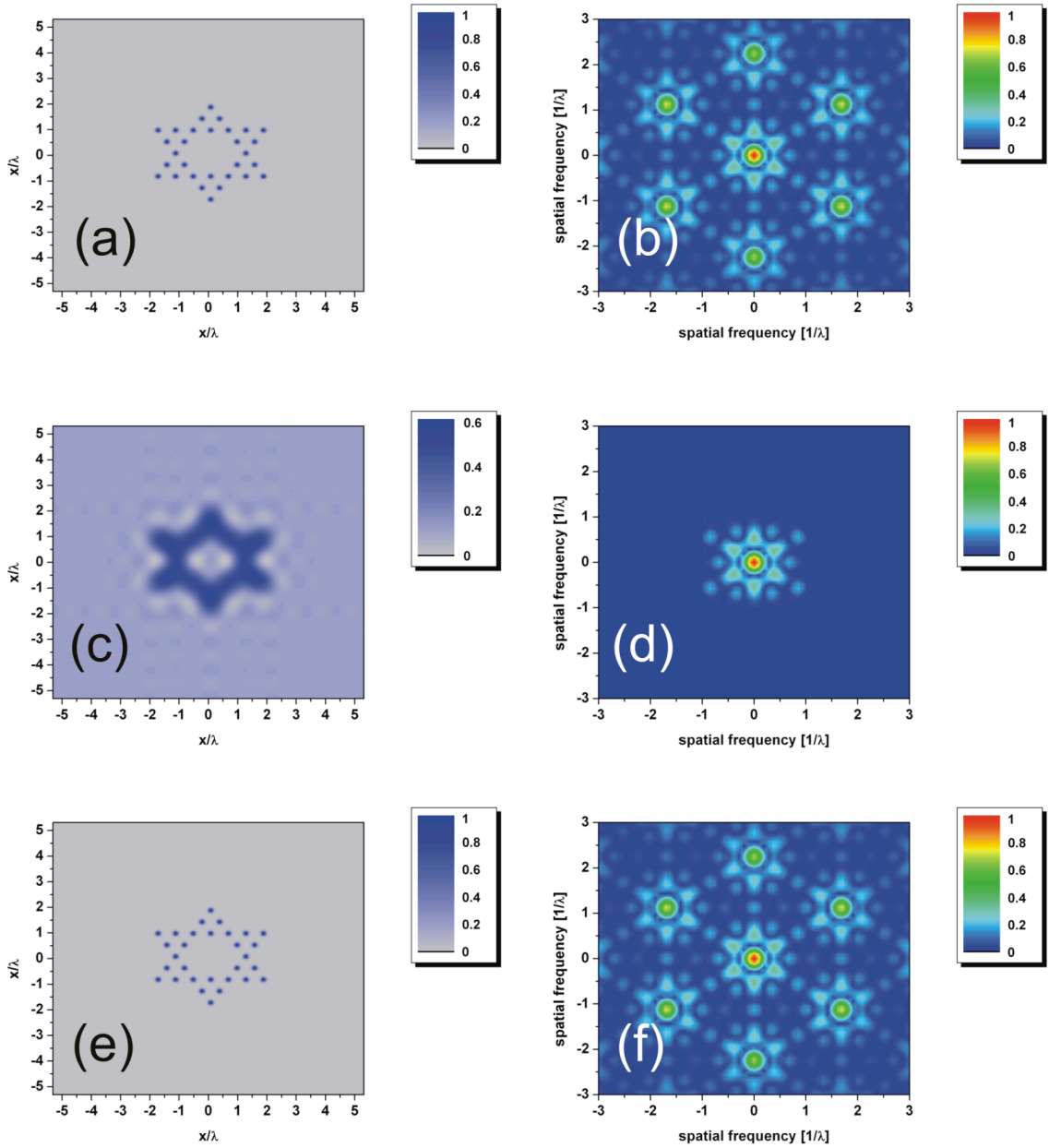